\documentclass[superscriptaddress,nofootinbib]{revtex4-2}

\usepackage{tikz}

\usetikzlibrary{decorations.pathmorphing}
\usetikzlibrary{decorations.markings}
\usetikzlibrary{positioning}
\usetikzlibrary{fadings,shadings,shapes}
\usetikzlibrary{patterns}
\definecolor{shinycol}{rgb}{0.75, 0.75, .85}
\definecolor{transcol}{rgb}{.85, .85, .9}
\usetikzlibrary{arrows}
\tikzset{cross/.style={cross out, draw, 
         minimum size=2*(#1-\pgflinewidth), 
         inner sep=0pt, outer sep=0pt}}
\usepackage{academicons}
\usepackage{hyperref}
\usepackage{orcidlink}
\definecolor{orcidlogocol}{HTML}{A6CE39}         
\newcommand{\msqrthz}{m/$\sqrt{\mathrm{Hz}}$ }
\newcommand\sqrthz{$\sqrt{\mathrm{Hz}}$}
\usepackage{acronym}
\usetikzlibrary{decorations.pathmorphing}
\usetikzlibrary{decorations.markings}
\usetikzlibrary{positioning}
\usetikzlibrary{fadings,shadings,shapes}
\usetikzlibrary{patterns}
\definecolor{shinycol}{rgb}{0.75, 0.75, .85}
\definecolor{transcol}{rgb}{.85, .85, .9}
\usetikzlibrary{arrows}
\usepackage{siunitx}
\usepackage{helvet}

\def\laser[(#1,#2),#3]{    
   \begin{scope}[transform canvas={rotate around={#3:(#1,#2)}}]
        \filldraw[fill=black!80](#1,#2-.5)rectangle(#1-1.4,#2-.25);
        \filldraw[fill=black!80](#1,#2+.5)rectangle(#1-1.4,#2+.25);
        \filldraw[fill=black](#1-1.5,#2-.4)rectangle(#1-1.4,#2+.4);
        \shade[top color=transparent!0, bottom color=red](#1-1.39,#2)rectangle(#1,#2+.25);
        \shade[top color=transparent!0, bottom color=red](#1-1.39,#2)rectangle(#1,#2+.25);     
        \shade[bottom color=transparent!0, top color=red](#1-1.39,#2+0.01)rectangle(#1,#2-.25); 
           \end{scope}
        \useasboundingbox (2,2) rectangle (-1,-1);
}

\def\mirrorPlane[(#1,#2),#3]{  
        \begin{scope}[transform canvas={rotate around={-#3:(#1,#2)}}]
        \filldraw[fill=black](#1+.15,#2-.4)rectangle(#1+.25,#2+.4);
        \draw[](#1,#2-.4)rectangle(#1+.15,#2+.4);
        \shade[top color=shinycol!100, bottom color=shinycol!100,middle color=shinycol!10](#1,#2-.4)rectangle(#1+.15,#2+.4);
        \end{scope}

        \useasboundingbox (2,2) rectangle (-1,-1);
}
\def\mirrorPlanePiezo[(#1,#2),#3]{  
        \begin{scope}[transform canvas={rotate around={-#3:(#1,#2)}}]
        \filldraw[fill=black](#1+.15,#2-.4)rectangle(#1+.25,#2+.4);
        \draw[](#1,#2-.4)rectangle(#1+.15,#2+.4);
        \shade[top color=shinycol!100, bottom color=shinycol!100,middle color=shinycol!10](#1,#2-.4)rectangle(#1+.15,#2+.4);
        \end{scope}
        \filldraw[fill=black](#1+.25,#2-.25)rectangle(#1+.45,#2+.25); 
        \useasboundingbox (2,2) rectangle (-1,-1);
}
\def\oscillator[(#1,#2),#3]{  
        \begin{scope}[transform canvas={rotate around={-#3:(#1,#2)}}]
       \draw[decoration={aspect=0.3, segment length=1, amplitude=1,coil},decorate] (#1+.15,#2+.2) -- (#1+.3,#2+.2);
       
       \draw[decoration={aspect=0.3, segment length=1, amplitude=1,coil},decorate] (#1+.15,#2-.2) -- (#1+.3,#2-.2);
        \draw[](#1,#2-.4)rectangle(#1+.15,#2+.4);
        \shade[top color=shinycol!100, bottom color=shinycol!100,middle color=shinycol!10](#1,#2-.4)rectangle(#1+.15,#2+.4);
        
        \filldraw[fill=black](#1+.3,#2-.4)rectangle(#1+.4,#2+.4);

        \end{scope}
        \useasboundingbox (2,2) rectangle (-1,-1);
}
\def\mirrorPlaneActuated[(#1,#2),#3]{  
        \begin{scope}[transform canvas={rotate around={-#3:(#1,#2)}}]
        \filldraw[fill=black](#1+.15,#2-.4)rectangle(#1+.25,#2+.4);
        \draw[](#1,#2-.4)rectangle(#1+.15,#2+.4);
        \shade[top color=shinycol!100, bottom color=shinycol!100,middle color=shinycol!10](#1,#2-.4)rectangle(#1+.15,#2+.4);
        
        \draw[thick](#1+.25,#2+.3)rectangle(#1+.45,#2+.1);
        \draw[thick](#1+.25,#2-.3)rectangle(#1+.45,#2-.1);
        \shade[top color=black!30, bottom color=black!30,middle color=black!90](#1+.25,#2+.3)rectangle(#1+.45,#2+.1);
        
        \shade[top color=black!30, bottom color=black!30,middle color=black!90](#1+.25,#2-.3)rectangle(#1+.45,#2-.1);
        \end{scope}
        
        \useasboundingbox (2,2) rectangle (-1,-1);
}
\def\mirrorConcave[(#1,#2),#3]{  
        \begin{scope}[transform canvas={rotate around={-#3:(#1,#2)}}]
        \filldraw[fill=black](#1+.15,#2-.4)rectangle(#1+.25,#2+.4);
        \draw[](#1-.09,#2-.4)arc(-20:20:1.15)--(#1+.15,#2+.4-.00814158435)--(#1+.15,#2+-.4)--(#1-.09,#2-.4);
            
                \shade[top color=shinycol!100, bottom color=shinycol!100,middle color=shinycol!10](#1-.09,#2-.4)arc(-20:20:1.15)--(#1+.15,#2+.4-.00814158435)--(#1+.15,#2+-.4)--(#1-.09,#2-.4); 
            
        \end{scope}
        
        \useasboundingbox (2,2) rectangle (-1,-1);
}
        
\def\mirrorConvex[(#1,#2),#3]{  
        \begin{scope}[transform canvas={rotate around={-#3:(#1,#2)}}]
        \filldraw[fill=black](#1+.15,#2-.4)rectangle(#1+.25,#2+.4);
        \draw[](#1+.09,#2-.5)arc(200:160:1.45)--(#1+.15,#2+.5-.00814158435)--(#1+.15,#2-.5)--(#1+.09,#2-.5);
            
                \shade[top color=shinycol!100, bottom color=shinycol!100,middle color=shinycol!10](#1+.09,#2-.5)arc(200:160:1.45)--(#1+.15,#2+.5-.00814158435)--(#1+.15,#2-.5); 
            
        \end{scope}
        
        \useasboundingbox (2,2) rectangle (-1,-1);
}

\def\polarisingBeamsplitter[(#1,#2),#3]{  
        \begin{scope}[transform canvas={rotate around={-45-#3:(#1,#2)}}]
        \draw[](#1-0.35355339059,#2)--(#1,#2-0.35355339059)--(#1+0.35355339059,#2)--(#1,#2+0.35355339059)--(#1-0.35355339059,#2); 
            
        \shade[left color=transcol!100, right color=transcol!100,middle color=transcol!10](#1-0.35355339059,#2)--(#1,#2-0.35355339059)--(#1+0.35355339059,#2)--(#1,#2+0.35355339059)--(#1-0.35355339059,#2);  
        
        \draw[](#1,#2-0.35355339059)--(#1,#2+0.35355339059);
        \end{scope}
        
        \useasboundingbox (1,1) rectangle (-1,-1);
}

\def\powerBeamsplitter[(#1,#2),#3]{  
        \begin{scope}[transform canvas={rotate around={-45-#3:(#1,#2)}}]
       \draw[](#1-.4,#2)rectangle(#1+.4,#2+.2); 
            
        \shade[top color=transcol!100, bottom color=transcol!10](#1-.4,#2)rectangle(#1+.4,#2+.2);

        \end{scope}
        
        \useasboundingbox (1,1) rectangle (-1,-1);
}
\def\lensVV[(#1,#2),#3]{  
        \begin{scope}[transform canvas={rotate around={#3:(#1,#2)}}]
        \draw[](#1,#2-.25*4/3)arc(200:160:.725*4/3)--(#1+0.05*4/3,#2+.25*4/3-.00814158435/2*4/3)arc(20:-20:.725*4/3)--(#1,#2-.25*4/3);
            
                \shade[top color=transcol!100, bottom color=transcol!100,middle color=transcol!10](#1,#2-.25*4/3)arc(200:160:.725*4/3)--(#1+0.05*4/3,#2+.25*4/3-.00814158435/2*4/3)arc(20:-20:.725*4/3)--(#1,#2-.25*4/3); 
            
        \end{scope}
}

\def\lensVP[(#1,#2),#3]{  
        \begin{scope}[transform canvas={rotate around={#3:(#1,#2)}}]
        \draw[](#1-0.05*2/3,#2-.5*2/3)arc(200:160:1.45*2/3)--(#1+0.05*2/3,#2+.5*2/3-.00814158435*2/3)--(#1+0.05*2/3,#2-.5*2/3)--(#1-0.05*2/3,#2-.5*2/3);
            
                \shade[top color=transcol!100, bottom color=transcol!100,middle color=transcol!10](#1-0.05*2/3,#2-.5*2/3)arc(200:160:1.45*2/3)--(#1+0.05*2/3,#2+.5*2/3-.00814158435*2/3)--(#1+0.05*2/3,#2-.5*2/3)--(#1-0.05*2/3,#2-.5*2/3); 
            
        \end{scope}
}
\def\lensVC[(#1,#2),#3]{  
        \begin{scope}[transform canvas={rotate around={#3:(#1,#2)}}]
        \draw[](#1-.1*2/3,#2-.5*2/3)arc(-20:20:1.45*2/3)--(#1-.1*2/3+0.2*2/3,#2+.5*2/3)--(#1+0.1*2/3,#2+.5*2/3)arc(20:-20:1.45*2/3)--(#1-.1*2/3,#2-.5*2/3);
            
                \shade[top color=transcol!100, bottom color=transcol!100,middle color=transcol!10](#1-.1*2/3,#2-.5*2/3)arc(-20:20:1.45*2/3)--(#1-.1*2/3+0.2*2/3,#2+.5*2/3)--(#1+0.1*2/3,#2+.5*2/3)arc(20:-20:1.45*2/3)--(#1-.1*2/3,#2-.5*2/3); 
            
        \end{scope}
}
\def\lensCC[(#1,#2),#3]{  
        \begin{scope}[transform canvas={rotate around={#3:(#1,#2)}}]
        \draw[](#1-0.05*2/3,#2-.5*2/3)arc(200:160:1.45*2/3)--(#1+0.05*2/3,#2+.5*2/3)arc(20:-20:1.45*2/3)--(#1-.05*2/3 ,#2-.5);
            
                \shade[top color=transcol!100, bottom color=transcol!100,middle color=transcol!10](#1-.05*2/3,#2-.5*2/3)arc(200:160:1.45*2/3)--(#1+0.05*2/3,#2+.5*2/3)arc(20:-20:1.45*2/3); 
            
        \end{scope}
}
\def\lensCP[(#1,#2),#3]{  
        \begin{scope}[transform canvas={rotate around={#3:(#1,#2)}}]
        \draw[](#1-0.05*2/3,#2-.5*2/3)arc(200:160:1.45*2/3)--(#1+0.1*2/3,#2+.5*2/3)--(#1+.1*2/3,#2-.5*2/3)--(#1-.05*2/3,#2-.5*2/3);
            
                \shade[top color=transcol!100, bottom color=transcol!100,middle color=transcol!10](#1-0.05*2/3,#2-.5*2/3)arc(200:160:1.45*2/3)--(#1+0.1*2/3,#2+.5*2/3)--(#1+.1*2/3,#2-.5*2/3)--(#1-.05*2/3,#2-.5*2/3); 
            
        \end{scope}
}
\def\waveplate[(#1,#2),#3,#4]{  
        \begin{scope}[transform canvas={rotate around={#3:(#1,#2)}}]
        \draw[](#1-0.05,#2-.3)rectangle(#1+.05,#2+.3);
            
                \shade[top color=#4!60, bottom color=#4!60,middle color=#4!10](#1-0.05,#2-.3)rectangle(#1+.05,#2+.3);

        \end{scope}
}
\def\faradayIsolator[(#1,#2),#3]{  
        \begin{scope}[transform canvas={rotate around={#3:(#1,#2)}}]
            \polarisingBeamsplitter[(#1-.65,#2),0];
            \polarisingBeamsplitter[(#1+.65,#2),0];
            \draw[thick ](#1-.4,#2-.4)rectangle(#1+.4,#2+.4);
            \shade[top color=black!30, bottom color=black!30,middle color=black!90](#1-.4,#2-.4)rectangle(#1+.4,#2+.4);
            
        \end{scope}
}
\def\faradayRotator[(#1,#2),#3]{  
        \begin{scope}[transform canvas={rotate around={#3:(#1,#2)}}]
            \polarisingBeamsplitter[(#1-.65,#2),0];
            \draw[ thick](#1-.4,#2-.4)rectangle(#1+.4,#2+.4);
            \shade[top color=black!30, bottom color=black!30,middle color=black!90](#1-.4,#2-.4)rectangle(#1+.4,#2+.4);
            
        \end{scope}
}
\def\beamDump[(#1,#2),#3,#4]{  
        \begin{scope}[transform canvas={rotate around={#3:(#1,#2)}}]
 
              \shade[top color=black!30, bottom color=black!30,middle color=black!90](#1,#2-.25*#4)rectangle(#1+.4*#4,#2+.25*#4);
            
            \fill[black](#1,#2-.3*#4)rectangle(#1+.1*#4,#2+.3*#4);
            
            \fill[black](#1+.15*#4,#2-.3*#4)rectangle(#1+.25*#4,#2+.3*#4);
            
            \fill[black](#1+.3*#4,#2-.3*#4)rectangle(#1+.5*#4,#2+.3*#4);
            
        \end{scope}
}

\def\EOM[(#1,#2),#3]{  
        \begin{scope}[transform canvas={rotate around={#3:(#1,#2)}}]
            \draw[](#1-.25,#2-.22)rectangle(#1+.25+.1,#2+.22);
            \shade[top color=yellow!0, bottom color=yellow!0,middle color=yellow!50](#1-.25,#2-.22)rectangle(#1+.25+.1,#2+.22);
            \filldraw[fill= brown](#1-.25,#2-.27)rectangle(#1+.25+.1,#2-.32);
            \filldraw[fill= brown](#1-.25,#2+.27)rectangle(#1+.25+.1,#2+.32);

        \end{scope}
}
\def\AOM[(#1,#2),#3]{  
        \begin{scope}[transform canvas={rotate around={#3:(#1,#2)}}]  
            \fill[black,rounded corners](#1-.2,#2+.3) rectangle(#1+.2+.1,#2);
            \draw[](#1-.3,#2-.22)rectangle(#1+.3+.1,#2+.22);
            \shade[top color=black!10, bottom color=black!10,middle color=black!80](#1-.3,#2-.22)rectangle(#1+.3+.1,#2+.22);
            \fill[ white,](#1-.15,#2-.17)rectangle(#1+.15+.1,#2+.22);
              
             \draw[blue!30](#1-.15,#2-.125)--(#1+.15+.1,#2-.125); 
             \draw[blue!30](#1-.15,#2-.075)--(#1+.15+.1,#2-.075); 
             \draw[blue!30](#1-.15,#2-.025)--(#1+.15+.1,#2-.025); 
            \draw[blue!30](#1-.15,#2+.025)--(#1+.15+.1,#2+.025);             \draw[blue!30](#1-.15,#2+.075)--(#1+.15+.1,#2+.075); \draw[blue!30](#1-.15,#2+.125)--(#1+.15+.1,#2+.125); 
            \draw[blue!30](#1-.15,#2+.175)--(#1+.15+.1,#2+.175); 
            \draw[thick](#1-.15,#2-.17)rectangle(#1+.15+.1,#2+.22);
            \fill[black](#1-.15,#2-.17)rectangle(#1+.15+.1,#2-.22);
        \end{scope}
}
\def\photodetector[(#1,#2),#3]{

        \begin{scope}[transform canvas={rotate around={#3:(#1,#2)}}]
            
            \shade[top color=red!100, bottom color=orange!50,middle color=orange!90](#1,#2+.2)--(#1,#2-.2)arc(-100:100:.19);
            \draw[](#1,#2+.2)--(#1,#2-.2)arc(-100:100:.19);
            \draw[very thick](#1,#2+.25)--(#1,#2-.25);

        \end{scope}

}
\def\QPD[(#1,#2),#3]{

        \begin{scope}[transform canvas={rotate around={#3:(#1,#2)}}]
            
            \shade[top color=red!90, bottom color=orange!50,middle color=orange!90](#1,#2+.2)--(#1,#2-.2)arc(-100:100:.197);
            \draw[](#1,#2+.2)--(#1,#2-.2)arc(-100:100:.197);
            \draw[thick](#1,#2+.25)--(#1,#2-.25);
            \draw[dashed](#1,#2+.2)--(#1+.5,#2+.2);
            \draw[dashed](#1,#2-.2)--(#1+.5,#2-.2);
            \draw[](#1+.5,#2)circle(.197);
            \shade[inner color=black!40, outer color=black!10](#1+.5,#2)circle(.195);
            
            \draw[](#1+.5,#2-.197)--(#1+.5,#2+.197);
            \draw[](#1+.5-.197,#2)--(#1+.5+.197,#2);

        \end{scope}

}

\def\genericAmp[(#1,#2),#3]{

        \begin{scope}[transform canvas={rotate around={#3:(#1,#2)}}]
            \draw[very thick](#1,#2+.4)--(#1+.7,#2)--(#1,#2-.4)--(#1,#2+.4)--(#1+.7,#2);

            \shade[left color=black!40, right color=black!40,middle color=black!0](#1,#2+.4)--(#1+.7,#2)--(#1,#2-.4)--(#1,#2+.4);

        \end{scope}

}
\def\mirroractConcave[(#1,#2),#3]{  
        \begin{scope}[transform canvas={rotate around={-#3:(#1,#2)}}]
        \filldraw[fill=black](#1+.15,#2-.4)rectangle(#1+.25,#2+.4);
        \draw[](#1-.09,#2-.4)arc(-20:20:1.15)--(#1+.15,#2+.4-.00814158435)--(#1+.15,#2+-.4)--(#1-.09,#2-.4);
            
                \shade[top color=shinycol!100, bottom color=shinycol!100,middle color=shinycol!10](#1-.09,#2-.4)arc(-20:20:1.15)--(#1+.15,#2+.4-.00814158435)--(#1+.15,#2+-.4)--(#1-.09,#2-.4); 
                    \draw[thick](#1+.25,#2+.3)rectangle(#1+.45,#2+.1);
        \draw[thick](#1+.25,#2-.3)rectangle(#1+.45,#2-.1);    
               
            \shade[top color=black!30, bottom color=black!30,middle color=black!90](#1+.25,#2-.3)rectangle(#1+.45,#2-.1);
           
            \shade[top color=black!30, bottom color=black!30,middle color=black!90](#1+.25,#2+.3)rectangle(#1+.45,#2+.1);
            \filldraw[fill=black](#1+.55,#2-.4)rectangle(#1+.45,#2+.4); 
        \end{scope}
        
        \useasboundingbox (2,2) rectangle (-1,-1);
}

\def\HVAmp[(#1,#2),#3]{

        \begin{scope}[transform canvas={rotate around={#3:(#1,#2)}}]
            \draw[very thick](#1,#2+.4)--(#1+.7,#2)--(#1,#2-.4)--(#1,#2+.4)--(#1+.7,#2);

            \shade[left color=black!40, right color=black!40,middle color=black!05](#1,#2+.4)--(#1+.7,#2)--(#1,#2-.4)--(#1,#2+.4);

        \end{scope}
        \begin{scope}[shift={({.25*cos(#3)},{.25*sin(#3)})}]
            \draw[very thick, -stealth]({#1-.015},#2+.2)--(#1-.1,#2)--(#1+.1,#2+0.05)--(#1-.05,#2-.25);
        \end{scope}    
}

\def\PID[(#1,#2),#3]{

        \begin{scope}[transform canvas={rotate around={#3:(#1,#2)}}]
            \draw[very thick](#1,#2+.4)--(#1+.7,#2)--(#1,#2-.4)--(#1,#2+.4)--(#1+.7,#2);

            \shade[left color=black!40, right color=black!40,middle color=black!05](#1,#2+.4)--(#1+.7,#2)--(#1,#2-.4)--(#1,#2+.4);

        \end{scope}
        \begin{scope}[shift={({.275*cos(#3)},{.275*sin(#3)})}]
               
                \node[font=\sffamily]at(#1,#2){\tiny{\bf{I}}};
                           
        \end{scope}   
        \begin{scope}[shift={({.12*cos(#3)},{.275*sin(#3)})}]
               
                    \node[]at(#1,#2-.175){\tiny{\bf{D}}}; \node[]at(#1,#2+.175){\tiny{\bf{P}}};

        \end{scope}   
}         
\def\filter[(#1,#2),#3]{

        \begin{scope}[transform canvas={rotate around={#3:(#1,#2)}}]
            \draw[very thick](#1-.3,#2-.3)rectangle(#1+.3,#2+.3);

            \shade[inner color=black!15, outer color=black!35](#1-.3,#2-.3)rectangle(#1+.3,#2+.3);
            
            \draw[very thick](#1-.2,#2-.2)--(#1-.1,#2-.2)--(#1+.1,#2+.2)--(#1+.2,#2+.2);

        \end{scope}

}
\def\mixerSummer[(#1,#2),#3]{

        \begin{scope}[transform canvas={rotate around={#3:(#1,#2)}}]
            \draw[very thick](#1,#2)circle(.3);

            \shade[inner color=black!35, outer color=black!05](#1,#2)circle(.3);
            
            \draw[very thick](#1-.15,#2-.15)--(#1+.15,#2+.15);    
            \draw[very thick](#1-.15,#2+.15)--(#1+.15,#2-.15);              
        
        \end{scope}

}
\def\mixerDifference[(#1,#2),#3]{

        \begin{scope}[transform canvas={rotate around={#3-45:(#1,#2)}}]
            \draw[very thick](#1,#2)circle(.3);

            \shade[inner color=black!35, outer color=black!05](#1,#2)circle(.3);
            
            \draw[very thick](#1-.15,#2-.15)--(#1+.15,#2+.15);

        \end{scope}

}
\def\sigGenerator[(#1,#2),#3]{
        \begin{scope}[transform canvas={rotate around={#3:(#1,#2)}}]
            \draw[very thick](#1,#2)circle(.3);

            \shade[inner color=black!35, outer color=black!0](#1,#2)circle(.3);
            
            \draw[very thick](#1-.15,#2)sin(#1-.075,#2+.15)cos(#1,#2)sin(#1+.075,#2-.15)cos(#1+.15,#2);  
        
        \end{scope}

}
\def\spectrumAnalyser[(#1,#2),#3]{
        \begin{scope}[transform canvas={rotate around={#3:(#1,#2)}}]
            \draw[rounded corners, thick](#1-.7,#2-.35)rectangle(#1+.65,#2+.4);
            
            \fill[rounded corners, black!10](#1+.05,#2-.1)rectangle(#1-.6,#2+.3);
            
            \fill[rounded corners=1pt, black!10](#1+.1,#2+.3)rectangle(#1+.3,#2+.2);
            \fill[rounded corners=1pt, black!10](#1+.1,#2+.05)rectangle(#1+.3,#2+.15);
            \fill[rounded corners=1pt, black!10](#1+.1,#2)rectangle(#1+.3,#2-.1);
            \fill[rounded corners=1pt, black!10](#1+.1,#2-.15)rectangle(#1+.3,#2-.25);
            
            \fill[rounded corners=1pt, black!10](#1+.35,#2+.3)rectangle(#1+.55,#2+.2);
            \fill[rounded corners=1pt, black!10](#1+.35,#2+.05)rectangle(#1+.55,#2+.15);
            \fill[rounded corners=1pt, black!10](#1+.35,#2)rectangle(#1+.55,#2-.1);
            \fill[rounded corners=1pt, black!10](#1+.35,#2-.15)rectangle(#1+.55,#2-.25);
            
            \fill[black!10](#1-.5,#2-.2)circle(.05);
            \fill[black!10](#1-.35,#2-.2)circle(.05);
            \fill[black!10](#1-.2,#2-.2)circle(.05);
            \fill[black!10](#1-.05,#2-.2)circle(.05);

        \end{scope}

}
\def\fibreCoupler[(#1,#2),#3]{
        \begin{scope}[transform canvas={rotate around={#3:(#1,#2)}}]
            \draw[thick](#1+.2,#2-.3)rectangle(#1,#2+.3);

            \draw[thick](#1-.2,#2-.1)rectangle(#1-.35-.2,#2+.1);
            \draw[thick](#1,#2-.4)rectangle(#1-.2,#2+.4);
            \shade[top color=black!20, bottom color=black!20,middle color=black!90](#1+.2,#2-.3)rectangle(#1,#2+.3);

            \shade[top color=black!20, bottom color=black!20,middle color=black!90](#1-.2,#2-.1)rectangle(#1-.35-.2,#2+.1);
        
            \shade[top color=black!30, bottom color=black!20,middle color=black!90](#1,#2-.4)rectangle(#1-.2,#2+.4);   
                        
        \end{scope}

}
\def\bandStop[(#1,#2),#3]{

        \begin{scope}[transform canvas={rotate around={#3:(#1,#2)}}]
            \draw[very thick](#1-.35,#2-.35)rectangle(#1+.35,#2+.35);

            \shade[inner color=black!35, outer color=black!05](#1-.35,#2-.35)rectangle(#1+.35,#2+.35);

        \end{scope}
        
        \draw[shift={(#1,#2+.1)},very thick,domain=-.2:.2, smooth, variable=\x]plot ({\x}, {-.2*exp(-\x*\x*500)});
 
}
\def\rampGenerator[(#1,#2),#3]{
        \begin{scope}[transform canvas={rotate around={#3:(#1,#2)}}]
            \draw[very thick](#1,#2)circle(.3);

            \shade[inner color=black!35, outer color=black!0](#1,#2)circle(.3);
            
            \draw[very thick](#1-.15,#2-.15)--(#1-.05,#2+.15)--(#1+.05,#2-.15)--(#1+.15,#2+.15);  
        
        \end{scope}

}
\def\highPass[(#1,#2),#3]{
        \begin{scope}[transform canvas={rotate around={#3:(#1,#2)}}]
            \draw[very thick](#1-.35,#2-.35)rectangle(#1+.35,#2+.35);

            \shade[inner color=black!35, outer color=black!05](#1-.35,#2-.35)rectangle(#1+.35,#2+.35);
            
            \draw[very thick](#1-.15,#2-.15)--(#1-.05,#2-.15)--(#1+.05,#2+.15)--(#1+.15,#2+.15);  
        
        \end{scope}

}
\def\lowPass[(#1,#2),#3]{
        \begin{scope}[transform canvas={rotate around={#3:(#1,#2)}}]
          \draw[very thick](#1-.35,#2-.35)rectangle(#1+.35,#2+.35);

            \shade[inner color=black!35, outer color=black!05](#1-.35,#2-.35)rectangle(#1+.35,#2+.35);
            
            \draw[very thick](#1-.2,#2+.15)--(#1-.05,#2+.15)--(#1+.05,#2-.15)--(#1+.2,#2-.15);  
        
        \end{scope}

}
\tikzset{cross/.style={cross out, draw, 
         minimum size=2*(#1-\pgflinewidth), 
         inner sep=0pt, outer sep=0pt}}

\usepackage{subfigure}
\usepackage{amssymb,mathrsfs}
\usepackage{siunitx}
\usepackage{helvet}
\usepackage{hyperref}
\hypersetup{
    colorlinks=true,
    breaklinks=true,
    citecolor=black,
    linkcolor=black,
    menucolor=black,
    urlcolor=black,
}
\acrodef{Q}[$Q$ factor]{Mechanical Quality factor}
\acrodef{ASD}{Amplitude Spectral Density}

\acrodef{IBS}{Ion Beam Sputtering}
\acrodef{MCCS}{Multi Channel Coherent Subtraction}

\definecolor{silver}{rgb}{0.75, 0.75, 0.75}
\usepackage{siunitx}
\definecolor{purple}{rgb}{.5,0.,1}
\begin{document}

\title{High Precision Inertial Sensors on a One Inch Diameter Optic}

\author{Jonathan J. Carter\orcidlink{0000-0001-8845-0900}}
\affiliation{ Max Planck Institute for Gravitational Physics (Albert Einstein Institute), Callinstr. 38, Hannover, Germany}
\affiliation{ Institute for Gravitational Physics of the Leibniz Universit\"at Hannover, Callinstr. 38, Hannover, Germany}

\author{Pascal Birckigt\orcidlink{0000-0001-8492-5964}}
\affiliation{Fraunhofer Institute for Applied Optics and Precision Engineering, Albert-Einstein-Str.~7, Jena, Germany}

\author{Oliver Gerberding\orcidlink{0000-0001-7740-2698}} 
\affiliation{Institut für Experimentalphysik, Universität Hamburg, Luruper Chaussee~149,
 Hamburg, Germany}

\author{Sina M. Koehlenbeck\orcidlink{0000-0002-3842-9051}}
\affiliation{ Max Planck Institute for Gravitational Physics (Albert Einstein Institute), Callinstr. 38, Hannover, Germany}
\affiliation{ Institute for Gravitational Physics of the Leibniz Universit\"at Hannover, Callinstr. 38, Hannover, Germany}

\acrodef{Q}[$Q$ factor]{Mechanical Quality factor}
\acrodef{LIGO}{the Laser Interferometer Gravitational Wave Observatories}
\acrodef{TED}{Thermoelastic Damping}
\acrodef{GS13}[GS-13]{Geotech Short Period Seismometer S-13}
\acrodef{L4C}[L-4C]{Sercel L-4C}
\acrodef{PDH}{Pound Drever Hall}
\acrodef{PZT}{Piezo Electric Stack}
\acrodef{UGF}{Unity Gain Frequency}
\acrodef{IIS}{Interferometric Inertial Sensor}
\acrodef{MEMS}{Micro-Electro-Mechanical Systems}
\date{\today }
\begin{abstract}
Compact, high-precision inertial sensors are needed to isolate many modern physics experiments from disturbances caused by seismic motion. We present a novel inertial sensor whose mechanical oscillator fits on a standard one-inch diameter optic. The oscillators achieve a Quality factor of over 600,000 and a resonance frequency of 50\,Hz, giving them a suspension thermal noise floor lower than all commercially available inertial sensors. The oscillator is combined with a Pound-Drever-Hall based readout scheme that achieves a displacement noise of 100\,f\msqrthz above 0.2\,Hz. We integrate the oscillator and readout to make two inertial sensors. Of order n$g$ performance is achieved in a broad band from 0.1\,Hz to 200\,Hz. Below 20\,Hz, the sensor presented here offers comparable performance to the best inertial sensors available today while being a fraction of the size. Above 20\,Hz, the sensor is, to the author's knowledge, the best demonstrated in the literature to date for a device of this style, with a self-noise floor of 0.1\,n$g$\sqrthz. The excellent performance of the sensors across the relevant seismic frequencies, vacuum compatibility, and compact size make it a prime candidate for integration into sophisticated seismic isolation schemes, such as those used by gravitational wave detectors. 

\end{abstract}

\maketitle

\section{Introduction}
Inertial sensors are used when we need to measure residual acceleration. They see wide use from spacecraft to mining exploration. However, one area where many might not consider their use is alongside precise physics experiments \cite{Matichard2015a,Richardson2020,Heijningen2023}, which often have to contend with seismic activity disturbing their sensitive measurements. A key example of this is gravitational wave detectors such as \ac{LIGO} \cite{Aasi2015}, which use a suite of inertial sensors to achieve the levels of seismic isolation required to operate \cite{Matichard2015a,Matichard2015b}. 
\par
High-precision inertial sensors are needed to isolate from seismic motion with active control schemes. To achieve a good performance, a traditional high-precision inertial sensor will use a large test mass, typically $\sim$kg \cite{Matichard2015a,Matichard2015b}. Large sensors are bulky, limiting the locations around the experiments where they can be deployed. Compact inertial sensors exist in the form of \ac{MEMS} sensors
 \cite{Yazdi1998, Zou2014, niu2007, trusov2013, Zotov2015}. These sensors typically have $\rm{\mu} \rm{g}$-ng of suspended test mass, but compensate for this with a high \ac{Q} oscillation. The low test mass prevents them from reaching anywhere near the same performance as the bulkier sensors due to increased suspension thermal noise. 
\par
Guzman et al. were the first to show a high Q oscillator with a gram scale mass \cite{Guzman2014, Gerberding2015}. The work showed that an oscillator design using two \SI{100}{\micro\meter} thick bridges can support larger test masses without degrading the \ac{Q}. Since then, various groups have begun evolution of the idea to adapt it for specific niches \cite{Carter2020a,Hines2020,Nelson2022,Kumanchik2023,Capistran2023,Hines2023}. There is a strong desire to produce a design suitable for use inside the seismic isolation system of a gravitational wave detector. To contribute meaningfully to the seismic isolation control scheme, the sensors must have sensitivity comparable to the current high-precision inertial sensors (for example, the \ac{L4C}). This means of order n$g$/$\sqrt{\mathrm{Hz}}$ sensitivity across all or part of the relevant control bandwidth (0.03-100\,Hz) \cite{Matichard2015a,Matichard2015b}.
\par
This paper explores the complete design of a novel inertial sensor suitable for this task. A mechanical oscillator is designed, manufactured, and tested to show that it is suitable for use in the inertial sensor. The oscillator, which encompasses all the mechanical parts of the complete sensor, is designed to fit on a 1" diameter optic. This mechanical oscillator is combined with a \ac{PDH} locked \cite{Drever1983} 5\,cm long optical resonator as a readout  of the test mass position. We show a design with excellent sensitivity in a compact housing. The inertial sensor presented here is the first with such a high \ac{Q} to be integrated with a high finesse optical resonator. At the same time, the oscillator achieves a m$Q$ product of 2 tonnes, the highest of the Author's Knowledge to date for a gram scale oscillator \cite{Hines2023,Kumanchik2023}. The self-noise of the device above 20\,Hz is also the lowest available at the time of publishing\cite{Cooper2018,Kirchhoff2020,Hines2023}.

\section{Inertial Sensor Design }
\label{sec:ISD}
A typical inertial sensor has two necessary parts. First, a means of suspending a test mass to remain inertially stable. Second, a means of reading out the distance between the suspended mass and the frame of reference. The relative motion of a suspended test mass, $\Delta X (f)$, to the absolute inertial motion of the system, $X_{\rm{g}}$, is given by  
\begin{equation}
    \frac{\Delta X (f)}{X_{\rm{g}}(f)}=\frac{-f^2}{f^2-{f_0}^2-\frac{i{f_0}^2}{Q}},
    \label{eqn:tfseismo}
\end{equation}
as a function of frequency, $f$, where $f_0$ is the natural 
frequency of oscillation of the fundamental mode, and $Q$ is the \acl{Q} defined as 
\begin{equation}
    Q=2\pi\frac{\rm{Energy \ Stored}}{\rm{Energy\ Dissipated\ per\ Oscillation\ Cycle}}.
    \label{eqn:Qwordydef}
\end{equation} 
\par
Inertial sensors should be designed to minimise all spurious noise terms that can limit their sensitivity. The noises can be broken down into those that cause the test mass position to change without external force and those that cause us to mismeasure the position of the test mass. 
\par
Suspension thermal noise is the primary source of noise that disturbs the test mass position. It originates from the thermally-driven excitations of the molecular degrees of freedom of the test mass coupling to test mass motion through the fluctuation-dissipation theorem. It is often the fundamental limit of a design. Equations defining the limits of this noise source have been well defined in several places \cite{saulson1994}.
Suspension thermal noise typically becomes a problem for inertial sensors that target a sensitive bandwidth below 100\,Hz\cite{Carter2020,Hines2023,Guzman2014}.
\par
How thermal noise scales depends on the damping mechanism. 
When the damping is related to internal mechanical behaviour, it usually depends on displacement. This is called structural damping. 
The acceleration noise from structural damping is given by  
\begin{equation}
	\tilde{A}(f)=\sqrt{\frac{8\pi k_{\rm{b}}Tf_0^2}{ mf Q}}
	\label{eqn:strucNoise}
\end{equation}
where $k_{\rm{b}}$ is the Boltzmann Constant and $T$ is the temperature.
The factors that make a low noise inertial sensor are already apparent. We need a high mass, low natural frequency, and high \ac{Q}. Large inertial sensors achieve low damping losses using large proof masses with soft suspensions. Gram-scale inertial sensors must compensate for this mass loss to achieve high precision by using high \acp{Q}, typically at least in the order of $10^4$ \cite{Guzman2014,Carter2020a,Hines2023}. 
To build a high-precision inertial sensor, we must build an oscillator capable of achieving these high \acp{Q} and integrate it with a readout method  capable of measuring the motion of the suspended test mass.
\subsection*{Designing Gram Scale Oscillator}
This work aims to produce an inertial sensor that could sit alongside other physics experiments, particularly optical ones. It was, therefore, decided to fit the entire mechanical part on the wafer of a standard optic, with 1" diameter and 1/4" thickness. As the overall aim of the work is to integrate such sensors inside gravitational wave observatories \cite{Aasi2015,Punturo2010}, the sensor's overall performance needed to be comparable to the high-performance sensors already used on site \cite{Matichard2015a,Matichard2015b}. With similar performance to the current sensors used there, but in a compact package, we can add to the control schemes of such isolation by having sensors whose measurements better reflect the isolation requirements of the control scheme.
\par
The key element controlling the mechanical behaviour of the oscillator is the design of the thin bridges connecting the suspended test mass to the outer frame, which we call the flexures. They must ensure a soft suspension to keep the mass inertially still, leading to a low $f_0$, be low loss to ensure an excellent thermal noise floor, be sufficiently resilient to not break under the weight of the test mass, and be designed to push higher order modes to higher frequencies.
\par
In a previous publication, the theoretical optimal design of gram scale resonators has been studied \cite{Carter2023}. The paper considers the case of a gram scale resonator and how to optimise the design to achieve the best thermal noise in a resonator that does not fracture under operation. We used the results from this paper to design the resonator used in this work. To allow for handling and shipping, a maximum load of 5\,$g$ was used to estimate the upper load an oscillator should survive. The estimates on material dissipation depth, surface and bulk loss from \cite{Carter2023} were also used. The target was to achieve a noise performance comparable to the state-of-the-art inertial sensors used at \ac{LIGO}. An initial study suggested that 50\,Hz was a good target for $f_0$ as it would require flexures that still fit in the constraint of a one-inch optic and were not too long to be made. The thermal noise was then optimised using the conditions in \cite{Carter2023} to get the optimal design for these constraints. The result was that the flexures must be at least 22\,mm long, \SI{180}{\micro\meter} thick, and a total of 4\,mm high (which in this case could be split into 4 1\,mm high flexures) while supporting 3\,g of test mass (including bonded mirror). The oscillators were made of Corning 7890-0F fused silica due to its high purity and low bulk loss \cite{Numata2002}. The estimated results suggested the \ac{Q} of the fundamental mode should be 500,000, limited by a combination of \ac{TED} \cite{lifshitz2000,Norris2005,Carter2023} and surface losses, \cite{Gretarsson1999}.
\par
\begin{figure}
    \centering
    \includegraphics[scale=.3]{ 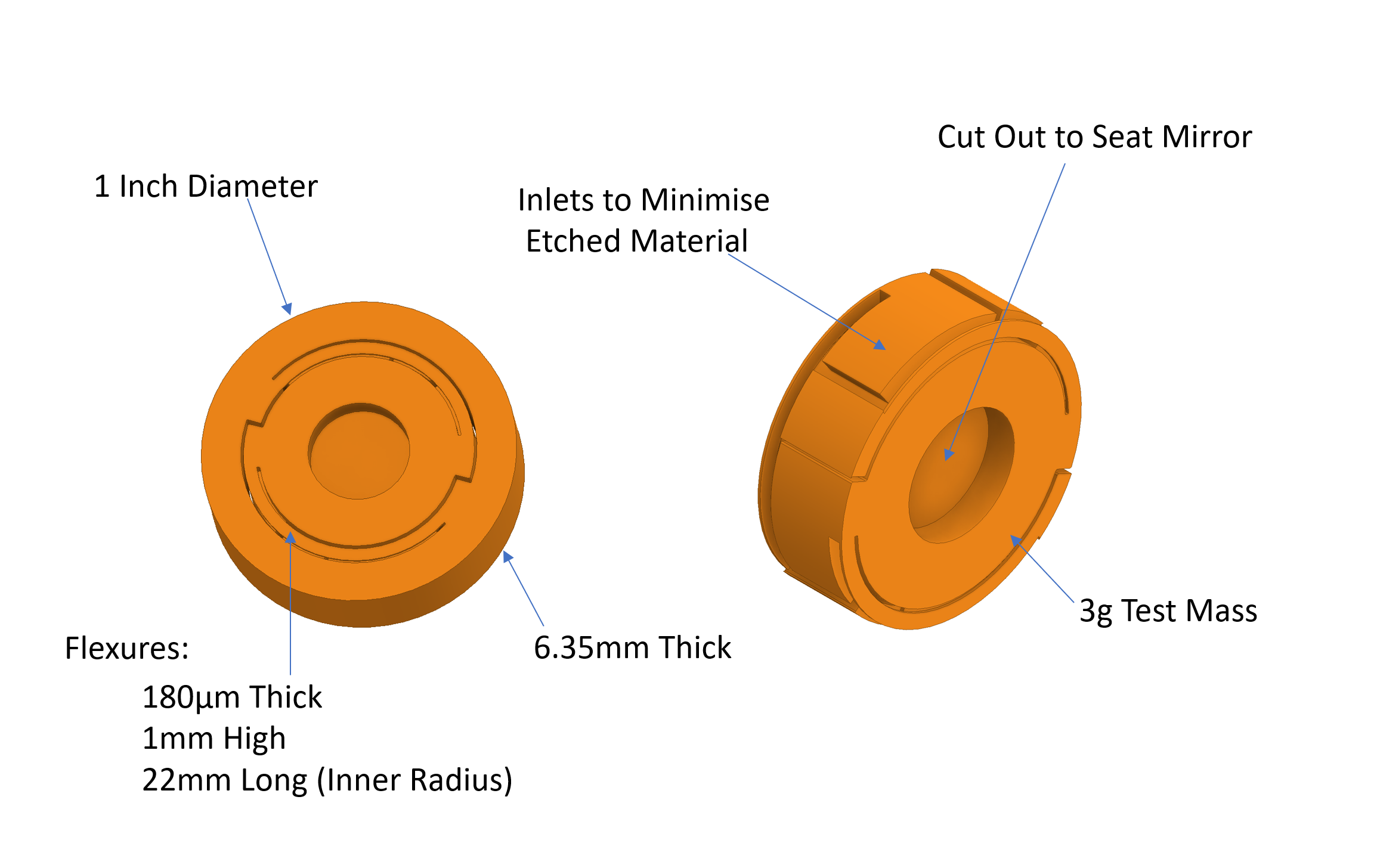}
    \caption{An annotated drawing of the key design features of the 50\,Hz oscillator. The flexure parameters were designed to make the oscillator withstand 5\,$g$ of acceleration while providing a $f_0$ of 50\,Hz . The second picture shows the inner structure. The material was carefully chosen to minimise the amount etched and prevent the flexures from contacting under maximum load. The inner radius of curvature was set at 0.1\,mm on all corners to distribute the stress localised in these areas.}
    \label{fig:50HAnno}
\end{figure}
\begin{figure}
    \centering
    \subfigure[50\,Hz Fundamental ``Corkscrew" Mode]{\includegraphics[width=.3\textwidth]{ 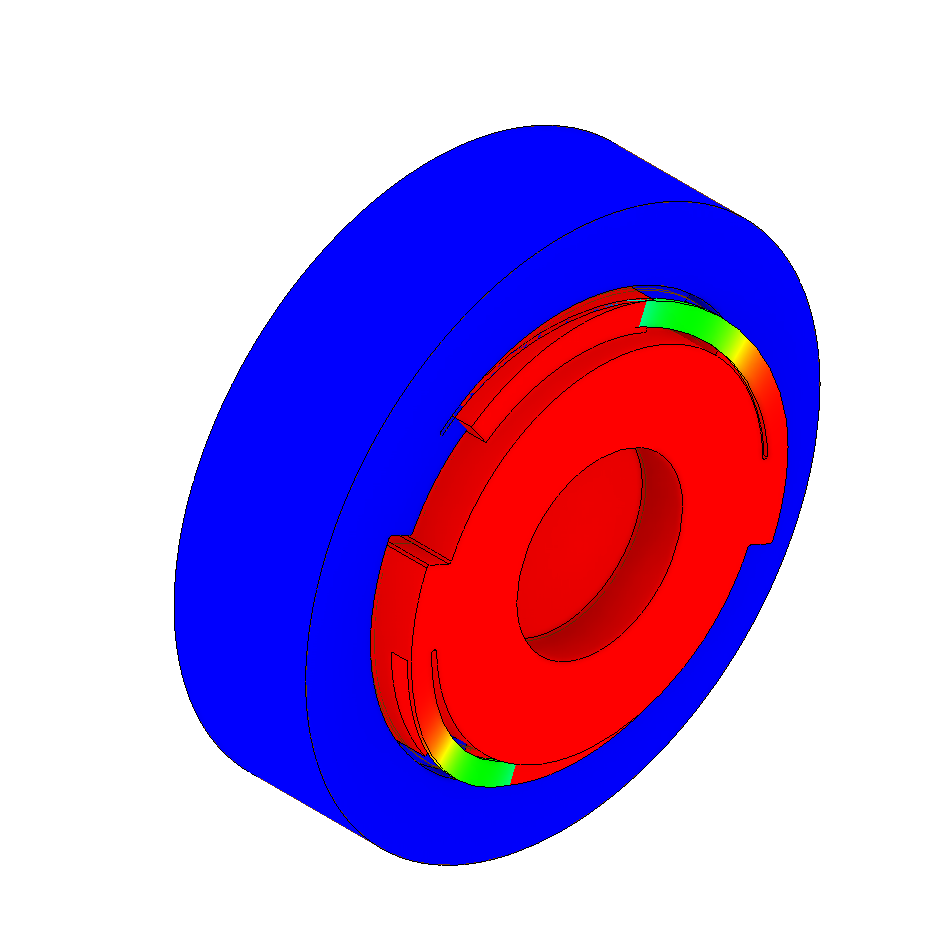}\hfill}
    \subfigure[209\,Hz 1st Order ``Tip-Tilt" Mode]{\includegraphics[width=.3\textwidth]{ 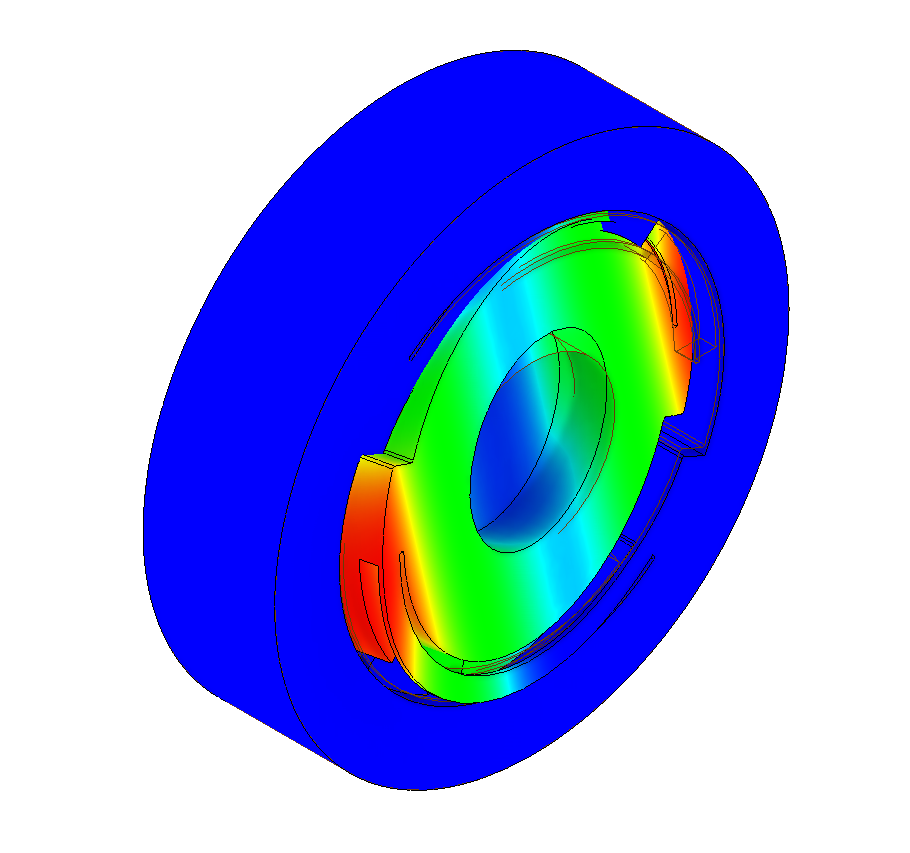}\hfill}
\caption{Fundamental and 1st order modes of the 50\,Hz oscillator. The colour scale indicates the moving parts during oscillation, with red indicating more motion and blue remaining stationary in the sensor's reference frame. The fundamental mode makes an in and out-of-plane oscillation similar to a corkscrew. This behaviour makes the resonator suitable for measuring motion in the direction of the optic plane. The 1st order mode defines the upper limit of the sensor's sensitive bandwidth. Measurements at frequencies below the fundamental frequency are possible using equation \ref{eqn:tfseismo}, but a reduced sensitivity is achieved the further below resonance we go.}
\label{fig:50HZmodes}
\end{figure}

With the flexure geometry decided, they must fit in the overall geometry of the piece. The final result of this is shown in Figure \ref{fig:50HAnno}. The aim is to maximise the separation between the fundamental and higher-order modes while minimising the amount of material removed. Previous resonators of a similar geometry have used parallelogram structures to force the motion to be constrained mainly to one dimension \cite{Guzman2014,Gerberding2015,Hines2023,Kumanchik2023}. Here, an alternating set of flexures in a disk creates a suspended test mass in the middle of the piece and a frame around the outside that can be clamped to the measurement frame of reference. The design results in a ``corkscrew" mode of oscillation as the fundamental mode, as shown in Figure \ref{fig:50HZmodes}. The mode travels directly in and out of the outer frame, meaning the sensor is largely only sensitive to oscillations from that degree of freedom. The next highest mode of oscillation is at 209\,Hz where several ``tip-tilt" modes occur. 
The oscillators required the attachment of a high-reflectivity coating. Previous experience with an \ac{IBS} of Niobola/Fused Silica resulted in a significant degradation of \ac{Q}. It was found that the attachment of small precoated mirrors with UV-cured Glue did not cause losses in \ac{Q}. To accommodate this, a small cutout was made on the oscillator to seat the mirror. The cutout would prevent a mass imbalance between the two sides of the oscillator and allow for easy alignment.
\section{Characterisation of Parts}
\label{sec:CoP}
The company FEMTOprint produced the resonators \cite{Bellouard2012} using a laser-assisted wet chemical etching technique. With the sensors produced, they were tested to verify the simulation's accuracy. Testing the sensors involved three steps: verifying the oscillator's mechanical performance, the optical readout performance, and the sensor's overall performance.
\subsection{Mechanical Behaviour}
Three parameters defined the mechanical behaviour of the oscillator: the suspended mass, $f_0$ and \ac{Q}. A test mass of 3.1\,g was measured.
$f_0$ and \ac{Q} were assessed by a ringdown experiment. A ringdown experiment involves exciting motion with a sharp impulse and tracking the decaying oscillation. The decay can be fit to an envelope of
\begin{equation}
    A(t)=A_0\, \rm{exp}\it \left({-\frac{\pi f_0 t}{Q}}\right),
    \label{eqn:ringdown}
\end{equation}
where $A$ is the position at time $t$, and $A_0$ is the amplitude of oscillation at time $t=0$. The specific implementation of this ringdown setup is described in detail in \cite{Carter2020}. The same heterodyne interferometer was used to track the motion of the test mass, with a piezo stack fixed to the base of the oscillator to excite motion. 
$f_0$ was estimated by taking the amplitude spectral density of the signal and finding the bin with the highest peak. For oscillators with high \ac{Q}.  MATLAB's nlinfit routine was used to extract the fit. The uncertainties quoted on measurements are reached from the covariance matrix on the fit. The first 10 minutes of data were not fitted to as higher order modes were also excited, and these were allowed to decay first to prevent spoiling the measurement. 
\par


The oscillators achieved a mean $f_0$ of 50.2\,Hz with a standard deviation of 0.5\,Hz, and a mean \ac{Q} of  637000, with a standard deviation of 18000. Each oscillator has a $f_0$ close to the target of 50\,Hz and well within the tolerance from manufacture, which gave an expected $f_0$ range of $\pm$8\,Hz, suggesting the quoted tolerance of average flexure thickness $\pm10$\,$\rm{\mu}$m was overestimated. Each oscillator achieved a \ac{Q} of over 600,000, higher than the initial estimate, indicating surface contamination was lower than expected. Regardless, these high \acp{Q} mean that the sensors would have a sufficiently low thermal noise such that the thermal noise floor of the oscillators is below that of many of the state-of-the-art inertial sensors available today across a large bandwidth, shown by Figure \ref{fig:readoutFixed}.

\subsection{Optical Readout of Test Masses}
\begin{figure}
    \centering
    \includegraphics[scale=0.75]{ 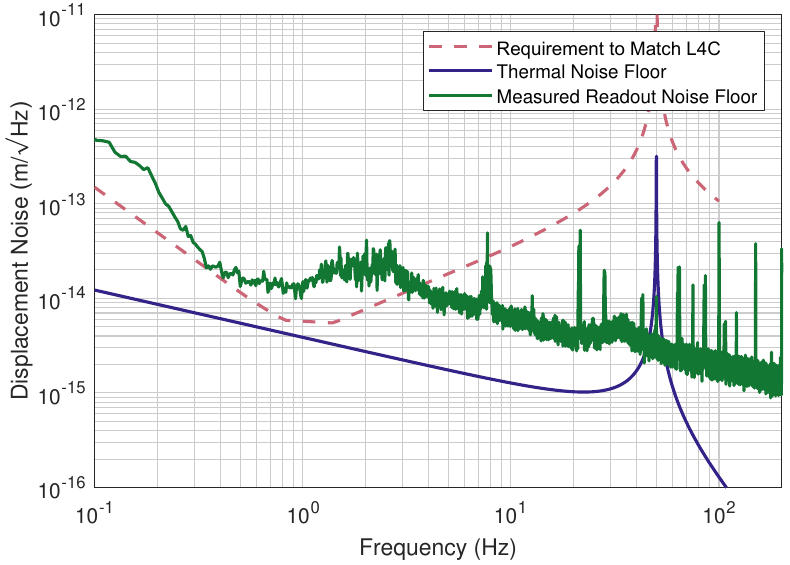}
    \caption{Assessment of the PDH locked optical resonators readout scheme when used with a 50\,Hz oscillator. The \ac{L4C} line corresponds to the requirement to match the sensors' performance. The thermal noise floor corresponds to the required read-out noise floor to match the thermal noise floor of the sensor. The green line is the measured noise floor of the readout scheme when a fixed mirror was used as an end mirror instead of an oscillator.   }
    \label{fig:readoutFixed}
\end{figure}

With an excellent thermal noise floor, the next step is to read out the position of the test mass with sufficient precision. There is a plethora of compact precision readout schemes, both commercial and in recent publications \cite{Gerberding2015,Cooper2018,Zhou2021,krause2012,Gerberding2015a,Isleif2019,Smetana2022,Yang2020,Smith2009,Zhang2022}. Figure \ref{fig:readoutFixed} shows the requirement for a readout scheme combined with these oscillators to match the thermal noise, with readout noise. The choice of readout for an inertial sensor will depend on the designer's experience with a specific technique, noise performance, and dynamic range of the readout compared to the expected motion of the test mass. For high \ac{Q} sensors in seismically active environments, the large relative test mass motion will force the usage of a readout scheme with high dynamic range, such as heterodyne interferometry. As these sensors were aimed to operate in a seismically quiet environment, the solution chosen here was to use a \acf{PDH} locked optical resonator readout \cite{Drever1983} due to the ability of such schemes to achieve a low noise floor. 
\par
\begin{figure}
 \centering
    \begin{tikzpicture}[>=latex,scale=1,decoration=snake,every node/.style={scale=1}]

        \def\vacnodex{5};       
        \def\vacnodey{-1};
        \def\vacHeight{8.4};
        \def\vacWidth{6};
        \def \fibreFBInterfacex{\vacnodex+3};
        \def \fibreFBInterfacey{\vacnodey+2};

        \def \OutofVacy{\fibreFBInterfacey-1};
        \def \cavLength{1.5};
        \def \preCavx{\vacnodex+\vacWidth/2-1};
        \def\lasnodey{\OutofVacy};
        \def\lasnodex{(\vacnodex-6.5};
        \def \FBfibreInterfacex{\lasnodex+3};
        \def \FBfibreInterfacey{\lasnodey};        
        \def \outBeamY{\vacnodey+4};
        \def \oisoOutY{\vacnodey+5.5}
        \def \oistOutY{\vacnodey+7}
        
        \filldraw[rounded corners, fill=black!10](\vacnodex,\vacnodey) rectangle(\vacnodex+\vacWidth,\vacnodey+\vacHeight);

        \draw[very thick, red](\lasnodex,\lasnodey)--( \FBfibreInterfacex,\FBfibreInterfacey);
        \draw[very thick, blue]( \FBfibreInterfacex,\FBfibreInterfacey)--(\fibreFBInterfacex,\fibreFBInterfacey-1)--(\fibreFBInterfacex,\fibreFBInterfacey);

        \draw[very thick, red](\fibreFBInterfacex,\fibreFBInterfacey)--(\fibreFBInterfacex,\outBeamY)--(\fibreFBInterfacex+1,\vacnodey+4)--(\vacnodex-.5,\outBeamY);
        \draw[black!0, dashed, very thick](\vacnodex,\outBeamY.4)--(\vacnodex,\fibreFBInterfacey+2.6);
           
        \draw[black, very thick,-triangle 45](\vacnodex-.5,\outBeamY)--(\lasnodex-2,\outBeamY)--(\lasnodex-2,\lasnodey)--(\lasnodex-1.5,\lasnodey);   
           
        \draw[very thick, red](\vacnodex+1,\outBeamY)--(\vacnodex-.5,\outBeamY);
        \polarisingBeamsplitter[(\fibreFBInterfacex,\outBeamY),0];
        \waveplate[(\fibreFBInterfacex+.5,\outBeamY),0,red]
        \laser[(\lasnodex,\lasnodey),0];    
        \faradayIsolator[(\vacnodex-5,\OutofVacy),0];
        \fibreCoupler[( \FBfibreInterfacex,\FBfibreInterfacey),180];            
        \EOM[(\vacnodex-2,\OutofVacy),0];

        \fibreCoupler[(\fibreFBInterfacex,\fibreFBInterfacey),90];

        \draw[line width=4pt, red!70](\fibreFBInterfacex+1,\vacnodey+4)--(\fibreFBInterfacex+1+\cavLength,\vacnodey+4);
        \mirrorConcave[(\fibreFBInterfacex+1,\vacnodey+4),180];
        \mirrorPlane[(\fibreFBInterfacex+1+\cavLength,\vacnodey+4),0]

        \photodetector[(\vacnodex-.5,\outBeamY),180];


        \draw[ thick,triangle 45-triangle 45](\vacnodex-2,\lasnodey+.4)--(\vacnodex-2,\outBeamY-.35);
        \draw[thick,-triangle 45](\vacnodex-2,1.5)--(\vacnodex-1.5,1.5)--(\vacnodex-1.5,\oisoOutY-.3);
         \draw[thick,-triangle 45](\vacnodex-2,1.5)--(\vacnodex-1,1.5)--(\vacnodex-1,\oistOutY-.3);
        \sigGenerator[(\vacnodex-2,1.5),0];

        \draw[](\vacnodex-3.6,1.5)node[right]{70\,MHz};
        \PID[(\vacnodex-4.5,\outBeamY),180];
        \HVAmp[(\vacnodex-6,\outBeamY),180];

        \draw[line width=4pt, red!70](\fibreFBInterfacex+1,\oisoOutY)--(\fibreFBInterfacex+1+\cavLength,\oisoOutY);
        \draw[very thick, blue](\fibreFBInterfacex-1,\lasnodey)--(\fibreFBInterfacex-1,\lasnodey+1);
        \draw[very thick, purple](\fibreFBInterfacex-1,\lasnodey+1.5)--(\fibreFBInterfacex-1,\oisoOutY)--(\fibreFBInterfacex+1,\oisoOutY)--(\vacnodex-.5,\oisoOutY);
        \draw[thick, black,-triangle 45](\vacnodex-.5,\oisoOutY)--(\vacnodex-4.5,\oisoOutY)--(\vacnodex-4.5,\oisoOutY-.75)--(\fibreFBInterfacex+.65,\oisoOutY-.75)--(\fibreFBInterfacex+.65,\oisoOutY-.3); \fibreCoupler[(\fibreFBInterfacex-1,\lasnodey+1.5),90];
        \polarisingBeamsplitter[(\fibreFBInterfacex-1,\oisoOutY),0];
        \mirroractConcave[(\fibreFBInterfacex+1,\oisoOutY),180];
        \oscillator[(\fibreFBInterfacex+1+\cavLength,\oisoOutY),0];
 
        \waveplate[(\fibreFBInterfacex,\oisoOutY),0,red];
        \photodetector[(\vacnodex-.5,\oisoOutY),180];
        \mixerSummer[(\vacnodex-1.5,\oisoOutY),180];
        \PID[(\vacnodex-2.5,\oisoOutY),180];
        \HVAmp[(\vacnodex-3.5,\oisoOutY-.75),0];       \draw[thick,dashed,-triangle 45](\vacnodex-1.5,\oisoOutY+.3)--(\vacnodex-1.5,\oisoOutY+.7)--(\vacnodex-6.4,\oisoOutY+.7);
        \draw[line width=4pt, red!70](\fibreFBInterfacex+1,\oistOutY)--(\fibreFBInterfacex+1+\cavLength,\oistOutY);
        \draw[very thick, blue](\fibreFBInterfacex-2,\lasnodey)--(\fibreFBInterfacex-2,\lasnodey+1.5);
       
        \draw[very thick, green](\fibreFBInterfacex-2,\lasnodey+1.5)--(\fibreFBInterfacex-2,\oistOutY)--(\fibreFBInterfacex+1,\oistOutY)--(\vacnodex-.2,\oistOutY);
        \fibreCoupler[(\fibreFBInterfacex-2,\lasnodey+2),90];
        \polarisingBeamsplitter[(\fibreFBInterfacex-2,\oistOutY),0]
        \draw[thick, black,-triangle 45](\vacnodex-.3,\oistOutY)--(\vacnodex-4.5,\oistOutY)--(\vacnodex-4.5,\oistOutY-.75)--(\fibreFBInterfacex+.65,\oistOutY-.75)--(\fibreFBInterfacex+.65,\oistOutY-.3); \fibreCoupler[(\fibreFBInterfacex-1,\lasnodey+1.5),90];
        
        \mirroractConcave[(\fibreFBInterfacex+1,\oistOutY),180];
        \oscillator[(\fibreFBInterfacex+1+\cavLength,\oistOutY),0];
      
        \waveplate[(\fibreFBInterfacex,\oistOutY),0,red];
        \photodetector[(\vacnodex-.2,\oistOutY),180];
        \mixerSummer[(\vacnodex-1,\oistOutY),180];
        \PID[(\vacnodex-2.5,\oistOutY),180];
        \HVAmp[(\vacnodex-3.5,\oistOutY-.75),0];
        
        \spectrumAnalyser[(\vacnodex-7,5),0];
        \draw[thick,dashed,-triangle 45](\vacnodex-5.5,\outBeamY)--(\vacnodex-5.5,3.8)--(\vacnodex-6.6,3.8)--(\vacnodex-6.6,4.7);
        \draw[thick,dashed,-triangle 45](\vacnodex-1.25,\outBeamY)--(\vacnodex-1.25,3)--(\vacnodex-3,3)--(\vacnodex-3,1.8)--(\vacnodex-7.5,1.8)--(\vacnodex-7.5,4.7);
  \mixerSummer[(\vacnodex-2,\outBeamY),180];  
        \draw[thick,dashed,-triangle 45](\vacnodex-4.5,\oisoOutY)--(\vacnodex-6,\oisoOutY)--(\vacnodex-6,\oisoOutY+.3)--(\vacnodex-6.4,\oisoOutY+.3);
        \draw[thick,dashed,-triangle 45](\vacnodex-4.5,\oistOutY)--(\vacnodex-6.6,\oistOutY)--(\vacnodex-6.6,\oistOutY-.6);
        \draw[thick,dashed,-triangle 45](\vacnodex-1,\oistOutY+.3)--(\vacnodex-1,\oistOutY+.7)--(\vacnodex-7.5,\oistOutY+.7)--(\vacnodex-7.5,\oistOutY-.6);
        \draw[](\fibreFBInterfacex+1+\cavLength*.5,\oisoOutY)node[above]{IIS 1};
        
        \draw[](\fibreFBInterfacex+1+\cavLength*.5,\oistOutY)node[above]{IIS 2};
        
        \draw[](\fibreFBInterfacex+1+\cavLength*.5,3)node[above]{Freq Ref};
         
         \node[left,align=right]at(-2.8,5){Control and\\ Data System};

        \node[below]at(\fibreFBInterfacex,\oistOutY-.3){QWP};
         
        \node[above,align=center]at(\fibreFBInterfacex+1,\oistOutY+.4){Piezo Driven \\Mirror};

        \node[right,align=left]at(\vacnodex+\vacWidth,\oistOutY){Mechanical\\ Oscillator};

        \node[left,align=left]at(\lasnodex+.2,\lasnodey+1){Laser Locked \\ to Freq Ref};

     \end{tikzpicture}  
     
     \caption{The layout of the huddle test measurement setup. Three sensors sat next to each other in vacuum. One frequency reference and two \ac{IIS}. Each Sensor had an identical input and readout setup. Each sensors had approximately 4\,mW of injected power, and where modulated with 70\,MHz sidebands, which were demodulated in the photodetector. The frequency reference cavity was use as a frequency reference for the laser. Meanwhile the two OIS were held in lock by using mechanical \acl{PZT} stacks mounted behind the in-coupling mirror to control the cavity length. }
    \label{fig:huddleLayout}
\end{figure}
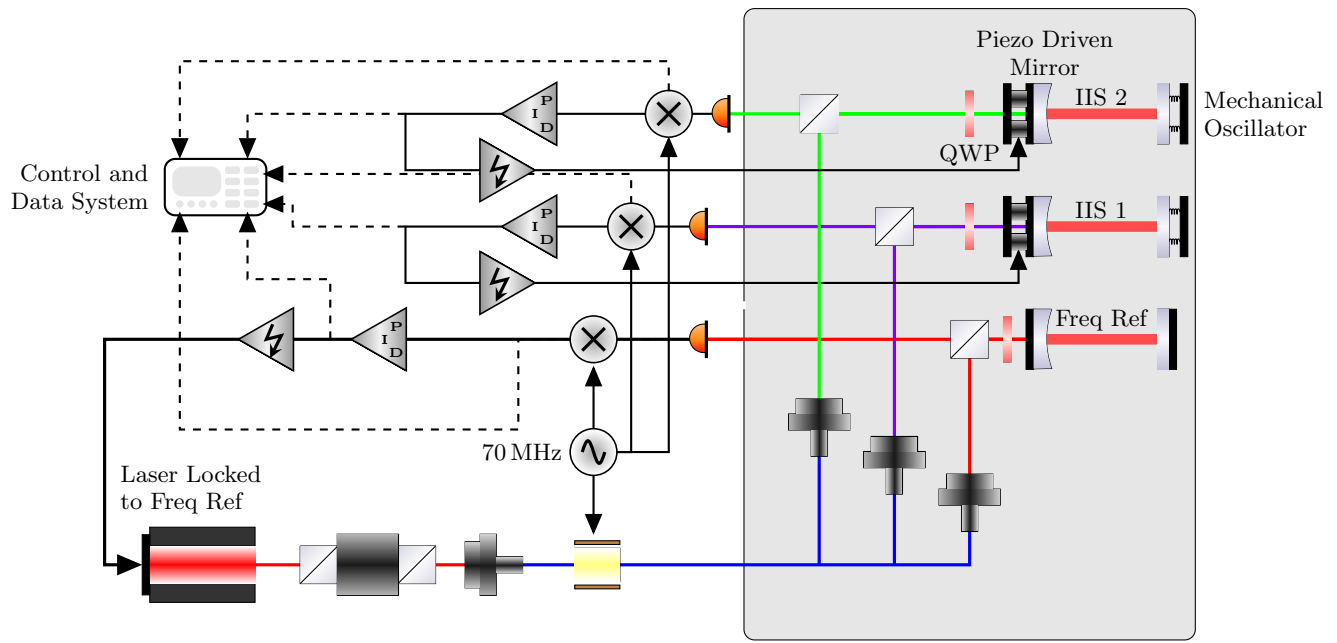

The schematic of this readout is shown in Figure \ref{fig:huddleLayout}. The oscillator is one end mirror of an optical resonator. The out-of-plane motion of the fundamental mode of an oscillator changes the length of the optical cavity. The length change causes a change in the demodulated \ac{PDH} signal. The cavity length must be controlled to keep the field in the cavity resonating. To do this, an active control loop was implemented in each resonator. 
A Piezo Electric Stack was mounted behind each in-coupling mirror of the \acf{IIS}. The cavity length could be tuned to stay on resonance by controlling the voltage on the Piezo Electric Stack.
The controls were implemented using an analogue servo. 
The control loops for the \ac{IIS} had \acl{UGF}
of $\sim$600\,Hz. A further readout scheme was used on a rigid cavity, which was used as a frequency reference. 
The frequency reference used the same optical design but was kept locked by shifting the laser's frequency. In addition, the mirrors were rigidly attached to an ultra-low expansion glass spacer to minimise thermal coupling to length changes in the cavity. The control loop of the frequency reference had two integrator stages, with a \ac{UGF} of 20\,kHz. The full parameters of the optical resonators can be found in Table \ref{tab:opparm}.
\begin{table}[t]
    \centering
    \begin{tabular}{|c|c|}
          \hline

      Parameter & Value \\
      \\
        \hline
     Laser Wavelength & 1064\,nm\\
     Incoupling Mirror Reflectivity & 98.5$\pm$0.4\%\\
     Oscillator Coating Reflectivity & 98.5$\pm$0.4\%\\
      Incoupling Mirror Radius of Curvature & 0.25\,m\\
      Oscillator Mirror Radius of Curvature & Flat\\
      Cavity Length & 5\,cm\\
      Free Spectral Range & 3\,GHz\\
     Line Width & 14\,MHz\\
      Finesse & 200\\
      Input Power & 4mW\\
      \hline
     \end{tabular}
   \caption{The parameters of the optical resonators used to read out test mass position in the inertial sensors.}
    \label{tab:opparm}
\end{table}
\par
\par
The performance of the optical readout was assessed by switching one of the oscillators in Figure \ref{fig:huddleLayout} with a rigid mirror of the same reflectivity. This allowed us to measure the sensor's readout noise floor, shown in Figure \ref{fig:readoutFixed} with the yellow line. Several other points of reference are added to this plot, with the requirements to meet an \ac{L4C}, and an oscillator's thermal noise when used with the 50\,Hz oscillator. The device will be limited by its readout across the whole sensitive bandwidth. The cavity's maximum length fluctuations from the piezo stack's thermal noise were estimated to be an order of magnitude smaller than the oscillator's and so not limiting. The limit on the readout sensitivity arises from limited gain on the frequency reference. Regardless, the readout scheme offers excellent sensitivity when compared to many others in recent literature \cite{Gerberding2015,Cooper2018,Zhou2021,krause2012,Gerberding2015a,Isleif2019,Smetana2022,Yang2020,Smith2009,Zhang2022,Heijningen2023,Heijningen2022}. 
Many of these schemes avoid the use of closed loop control, giving them dynamic range benefits; however, by using closed loop control we are able to achieve sufficient sensitivity to integrate a high-frequency oscillator, which allows for a small device, that is easier to handle. 
In addition, it is sufficient to achieve comparable performance to a \ac{L4C} across most of the sensitive bandwidth.

\section{Evaluation of Sensor Performance}
\label{sec:EoSP}
With the sensors designed and assembled, we wished to evaluate the sensor's overall performance. To do this, a huddle test was performed \cite{Kirchhoff2017}. Here, multiple sensors are clustered together with their sensitive axis aligned. The sensors can then subtract the coherent parts of their signal using a \ac{MCCS} routine  \cite{Allen1999}. Doing so allows us to measure the true noise of the each of sensors, even when the measurements are polluted from residual vibrations disturbing the measurement. The whole setup in Figure \ref{fig:huddleLayout} was used to perform this experiment where two \ac{IIS} were set up, along with a frequency reference. Ground-mounted L4Cs were also used for additional subtraction. 
\par
\begin{figure}
    \centering
    \includegraphics[scale=0.75]{ 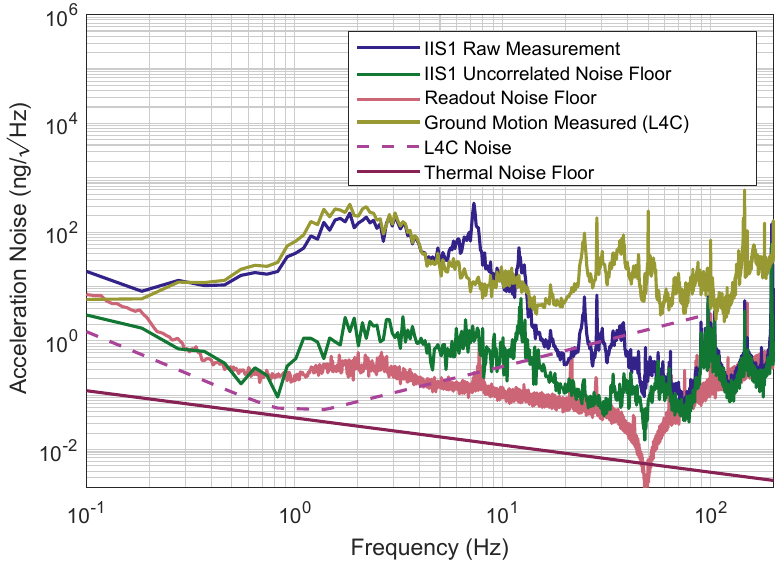}
    \caption{The results of the huddle test of two IIS with a frequency reference and ground mounted \ac{L4C}s as a comparison. The readout noise floor is the readout measured in Figure \ref{fig:readoutFixed} projected into acceleration units. The yellow and blue curves show the raw measured signal from a geophone mounted on top of the vacuum tank and the IIS, respectively. These have good agreement below 7\ Hz, showing the sensors are well calibrated. The IIS sits on a passive isolation stage with a resonance at 7\ Hz, which causes the discrepancy in raw signals above this frequency. In green, the noise of the IIS signal after all coherent signals have been subtracted is shown The self-noise of the IIS is limited by readout noise below 1\,Hz and above 20\,Hz. Between these values, the sensor's measurement is likely limited by motion differences between the two sensors. The self noises of an \ac{L4C} and the suspension thermal noise floor are also plotted as a reference.}
    \label{fig:finalres}
\end{figure}
The experiment was set up, and measurements were taken overnight and at weekends to find seismically quiet data stretches. The measured noise from \ac{IIS}1 is shown in Figure \ref{fig:finalres}. The pre-subtraction measurement agrees well with the noise of the \ac{L4C} below 7\,Hz as we expect, showing the sensor is well calibrated. The result of the coherent subtraction is shown in green in this Figure. The sensor is limited by the displacement noise above 20\,Hz and below 1\,Hz. Between these two limits, the sensor is neither limited by readout nor thermal noise. As the shape is similar to that of the pre-subtraction signal, likely, the measured noise is still residual motion differences between the two sensors. Likely, this is due to slight differences in the sensitive axis of the two \ac{IIS}. The sensors must be tested in a better-isolated environment to verify this. The device's sensitivity is, therefore, still limited by the displacement sensitivity. Below 1\,Hz the sensors perform better than the displacement sensitivity, suggesting a coherent noise source is dominating the readout. This is not seen in the pure displacement readout, as only a single resonator was tested. At higher frequencies, the noise is shown to be slightly lower than the rigid case because the rigid cavity had a slightly lower finesse and, therefore, greater noise.
\par
The \ac{IIS} achieves a sub-n$g$ performance across most of the control bandwidth of gravitational wave detectors (0.3-100\,Hz), with the exception of a few peaks of measured noise. Above 20\,Hz the sensor is one of the best available. The performance across a large band is entirely limited by the readout method. Work is ongoing to develop readout schemes that could achieve even more sensitive noise floors \cite{Heijningen2023}.
\par
Although the sensor does not quite reach the performance of larger inertial sensors at low frequencies, it has several advantages over any device available today. Its overall compact footprint means it can easily be deployed in many difficult-to-reach locations without unbalancing them. A fibre optic can be used to inject the laser beam into the sensor, meaning the compact sensor can be deployed far away from any data processing. The whole device is vacuum compatible and so can easily be deployed alongside many physics experiments without the usual need for a vault. Furthermore, as the device can operate in any orientation, multiple devices can be used to create 3D sensing in any arbitrary orientation.
\section{Conclusions}
By compacting down the performance of an inertial sensor onto a 1" optic, we show a widely deployable sensor that can be used alongside a variety of physics experiments. By integrating the mechanical performance onto such a small device, groups can fit these sensors close to their sensitive experiments. Many groups interested in using these sensors would already have the expertise and equipment necessary to read out the test mass position with sufficient sensitivity to use such an inertial sensor in these experiments. 
\par
Overall, the two required pieces for an inertial sensor, a suspension and displacement sensor, are working, achieving outstanding performance for a sensor of such a small size. The performance of the sensors has been validated to be near its expected noise floor using a simple laboratory setup.
\par
Ultimately, adopting such sensors alongside larger physics experiments will allow us to isolate them from the scourge of residual vibration disturbances, allowing us better to probe many of the open questions in physics today. 
\section*{Acknowledgements}
The authors would like to acknowledge insightful discussion with Harald Lück and Gerhard Heinzel.
\par
JJC, PB, and SMK acknowledge funding in the framework of the Max-Planck-Fraunhofer Kooperationsprojekt ``Glass Technologies for the Einstein Telescope (GT4ET)".
\par
OG was funded by the Deutsche Forschungsgemeinschaft (DFG, German Research Foundation) under Germany's Excellence Strategy—EXC 2121 “Quantum Universe”—390833306.

\end{document}